\documentclass[twocolumn]{jpsj2}
\def\runtitle{
Dispersive Gap Mode of Phonons in Anisotropic Superconductors
}

\def\runauthor{
Masanori \textsc{Ichioka}  and Kazushige \textsc{Machida}
}

\title{
Dispersive Gap Mode of Phonons in Anisotropic Superconductors
}

\author{
Masanori \textsc{Ichioka}\thanks{E-mail address: oka@mp.okayama-u.ac.jp}
and Kazushige \textsc{Machida}
}

\inst{Department of Physics, Okayama University, Okayama 700-8530}

\recdate{
April 4, 2003
}

\abst{
We estimate the effect of the superconducting gap anisotropy 
in the dispersive gap mode of phonons, which is observed 
by the neutron scattering on borocarbide superconductors. 
We numerically analyze the phonon spectrum considering 
the electron-phonon coupling, 
and examine contributions coming from the gap suppression and 
the sign change of the pairing function on the Fermi surface. 
When the sign of the pairing function is changed by the nesting 
translation, the gap mode does not appear. 
We also discuss the suppression of the phonon softening of the 
Kohn anomaly due to the onset of superconductivity.  
We demonstrate that observation of the gap dispersive mode is useful 
for sorting out the underlying superconducting pairing function. 
}

\kword{
gap mode in phonons, 
anisotropic superconducting gap, 
electron-phonon coupling, 
borocarbide.
}

\begin{document}

%

\sloppy
\maketitle


\section{Introduction}
\label{sec:introduction}

In the materials with strong electron-phonon interactions, 
we can obtain the information of the electronic state 
by investigating phonon properties. 
The Kohn anomaly of the phonon dispersion comes from 
the nesting features of the band structure in simple metals 
via electron-phonon interaction.~\cite{Kohn} 
It is known that the gap mode of phonons appears at $\omega=2 \Delta$ 
in the superconducting state with a gap $\Delta$. 
In 2H-NbSe$_2$, the Raman scattering~\cite{Sooryakumar} observes the gap 
mode around zero phonon wave vector ${\mib q}=0$.
The gap mode at ${\mib q} \ne 0$ is reported by the neutron scattering 
on YNi$_2$B$_2$C.~\cite{Kawano,KawanoC,KawanoB}  
There, the phonon dispersion shows the Kohn anomaly at the nesting 
wave vector ${\mib Q}\sim \pi(0.55,0,0)$,\cite{Dugdale} 
and the new peak of the dispersive gap mode appears below the original 
phonon dispersion around ${\mib q}={\mib Q}$. 
The neutron scattering on LuNi$_2$B$_2$C reports that the peak 
energy at the phonon dispersion is shifted to lower energy and that the 
peak intensity is enhanced below the superconducting transition 
temperature $T_{\rm c}$.~\cite{Stassis} 
Theoretically, the gap mode is obtained by considering the contribution 
of the electron-phonon coupling in the phonon 
selfenergy.\cite{Balseiro,Littlewood,Allen,Kee}
However, previous theoretical works have investigated 
in the isotropic $s$-wave superconductivity case, 
and they have not considered the dispersive behavior of the gap mode.  

Band calculations suggest that borocarbide superconductors are 
conventional BCS superconductors with the pairing interaction mediated 
by the electron-phonon interaction.~\cite{Pickett,Mattheiss,Coehoorm,Lee}
However, the direct observation of the superconducting gap of 
YNi$_2$B$_2$C by photoemission spectroscopy 
reports that the gap has large anisotropy and that 
the minimum of the gap seems to be zero as if it has point or line 
nodes.\cite{Yokoya}  
And the impurity effects suggest that they are extreme 
anisotropic $s$-wave superconductors.
This gap structure is also supported by the specific heat 
measurement.\cite{Nohara1,Nohara2} 
Thermal conductivity experiment by rotating magnetic field 
suggests that the superconducting gap has point nodes.\cite{Izawa} 
Therefore, it is important to study how the superconducting 
gap anisotropy affects the dispersive gap mode of phonons. 

In this paper, we investigate the effect of the gap anisotropy on 
the gap mode of phonons by calculating the dispersion and intensity 
of the gap mode around ${\mib q}={\mib Q}$. 
As the gap anisotropy effect, we examine the case when 
the superconducting gap is suppressed around the node points, 
and the case when the order parameter changes the sign by the 
nesting translation on the Fermi surface as in the $d$-wave pairing. 
The phonon properties are studied by varying $2\Delta$ 
from the weaker case to the stronger case relative to the phonon frequency. 
And we discuss the experimental data observed in YNi$_2$B$_2$C and 
LuNi$_2$B$_2$C. 
To obtain the renormalized phonon Green's function 
in the random phase approximation (RPA), 
we calculate the dynamical charge susceptibility in anisotropic 
superconductors. 
In the calculation, we use the electron dispersion characterized by 
the nesting vector ${\mib Q}\sim \pi(0.55,0,0)$, which reproduce the Kohn 
anomaly. 
When the Kohn anomaly is strong, the phonon is softened at 
${\mib q}\sim{\mib Q}$, indicating the transition to the charge density 
wave (CDW) state. This softening is suppressed by the onset of  
superconductivity. 

After explaining our formulation in \S\ref{sec:formulation}, 
we present our results of the phonon spectral function 
and discuss the effect of anisotropic superconductivity 
in \S\ref{sec:result}.  
The suppression of the phonon softening is discussed in \S\ref{sec:cdw}. 
The last section is devoted to summary and discussions.

\section{Formulation}
\label{sec:formulation}

\begin{figure}[tbh]
\begin{center}
\includegraphics{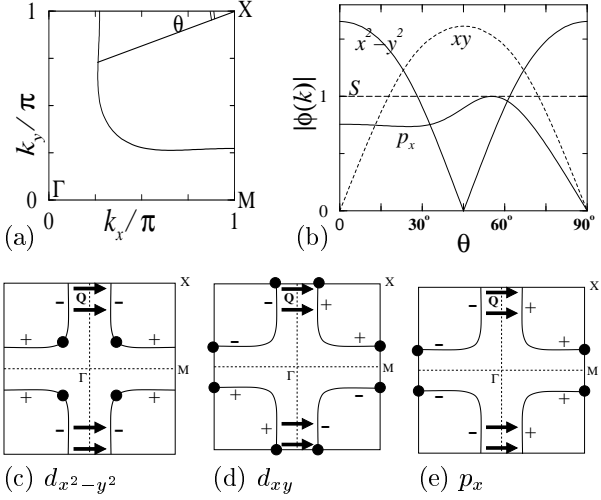}
\end{center} 
\caption{\label{fig:FS}
(a) Fermi surface for $t=1$, $t'=-0.6$, $t''=0.1$ and $\mu=-1.05$ 
in momentum region $0 \le k_x \le \pi$ and  $0 \le k_y\le \pi$. 
(b) The gap amplitude of the pairing functions along the Fermi surface. 
We plot $|\phi_s({\mib k})|$, 
$|\phi_{d_{x^2-y^2}}({\mib k})|=|\phi_{s_{x^2-y^2}}({\mib k})|$, 
$|\phi_{d_{xy}}({\mib k})|=|\phi_{s_{xy}}({\mib k})|$, and 
$|\phi_{p_{x}}({\mib k})|$ as a function of $\theta$, 
which is the angle centered at the X point as shown in (a). 
The lower panels schematically present the sign of the pairing function 
along the Fermi surface for the $d_{x^2-y^2}$-wave (c), 
the $d_{xy}$-wave (d) and the $p_{x}$-wave (e) cases 
in the region $-\pi \le k_x \le \pi$ and $-\pi \le k_y \le \pi$.  
The solid circles on the Fermi surface present the node points 
of the pairing functions. 
Arrows show the nesting vector ${\mib Q}$. 
}
\end{figure}

As the Fermi surface shape of borocarbide superconductors 
is complicated,\cite{Dugdale,Lee} 
we simplify the dominant large Fermi surface to the simple 
two-dimensional Fermi surface which 
reproduces the nesting property of nesting vector 
${\mib Q}=(Q_x,0,0) \sim \pi(0.55,0,0)$ as in the borocarbides. 
Then, we use the following dispersion of the electrons, 
\begin{eqnarray} 
K_{\mib k}&=& -2t(\cos k_x +\cos k_y ) -4t' \cos k_x \cos k_y 
\nonumber \\ && 
-2t''(\cos 2 k_x +\cos 2 k_y ) -\mu, 
\end{eqnarray}
with the transfer between nearest ($t$), second nearest ($t'$) and third 
nearest ($t''$) neighbor sites in two-dimensional square lattice 
of the basal plane in tetragonal crystal. 
The lattice constant is taken to unity.  
Here, we set $t=1$, $t'=-0.6$, $t''=0.1$ 
and the chemical potential $\mu=-1.05$. 
The energy and temperature are scaled by $t$ throughout this paper. 
The Fermi surface obtained by $K_{\mib k}=0$ is shown in 
Fig. \ref{fig:FS}(a). 
Since it has flat surface at $k_x \sim \pm 0.275 \pi$, 
it gives good nesting for ${\mib Q}\sim \pi(0.55,0,0)$. 

As for the pairing function $\phi({\mib k})$, we use 
$\phi_s({\mib k})=1$ for isotropic $s$-wave pairing,  
$\phi_{d_{x^2-y^2}}({\mib k})=\cos k_x -\cos k_y$ for $d_{x^2-y^2}$-wave, 
and $\phi_{d_{xy}}({\mib k})=2\sin k_x \sin k_y$ for $d_{xy}$-wave. 
The gap amplitude along the Fermi surface is shown in Fig. \ref{fig:FS}(b). 
The flat part of the Fermi surface related to the nesting is ranged for 
$0 \le \theta < 30^\circ$. 
At $\theta=0$, $|\phi_{d_{x^2-y^2}}({\mib k})|$ has large gap, and 
$|\phi_{d_{xy}}({\mib k})|=0$ due to the node structure. 
With increasing $\theta$, $|\phi_{d_{x^2-y^2}}({\mib k})|$ 
($|\phi_{d_{xy}}({\mib k})|$) is decreased (increased). 
These gap dependences on $\theta$ are averaged along the flat Fermi surface 
when they contribute to the phonon properties. 

It is also important to see the sign change of the pairing function 
through the nesting translation ${\mib Q}$. 
In the $d_{x^2-y^2}$-wave case, there are nodes along the diagonal 
lines $k_y=\pm k_x$. 
In the $d_{xy}$-wave case, nodes are located along the horizontal lines 
$k_y=0$, $\pm\pi$ and the vertical lines $k_x=0$, $\pm\pi$. 
The sign of $\phi_{d_{x^2-y^2}}({\mib k})$ is the same at the 
both ends connected by the nesting ${\mib Q}$, 
as schematically shown in Fig. \ref{fig:FS}(c).  
But, the sign of $\phi_{d_{xy}}({\mib k})$ changes by the nesting ${\mib Q}$, 
as shown in Fig. \ref{fig:FS}(d).  

For extreme anisotropic $s$-wave superconductors, we use 
$\phi_{s_{x^2-y^2}}({\mib k})=|\phi_{d_{x^2-y^2}}({\mib k})|$ and  
$\phi_{s_{xy}}({\mib k})=|\phi_{d_{xy}}({\mib k})|$. 
We call them as the $s_{x^2-y^2}$ case and the $s_{xy}$ case, respectively. 
They have the same superconducting gap as those of the $d_{x^2-y^2}$-wave 
and the $d_{xy}$-wave cases. 
But they do not change the sign of the pairing function 
under $\pi/2$-rotation. 
Then, the pairing function does not change the sign by the nesting ${\mib Q}$. 
We also consider the $p_x$-wave pairing case 
$\phi_{p_x}({\mib k})=\sin k_x $, while it is not for the 
pairing of borocarbide superconductors. 
The pairing function $\phi_{p_x}({\mib k})$ has nodes 
along the vertical lines $k_x=0$, $\pm\pi$. 
As shown in Fig. \ref{fig:FS}(b), $\phi_{p_x}({\mib k})$ has large gap at 
the flat Fermi surface $0 \le \theta < 30^\circ$. 
But, $\phi_{p_x}({\mib k})$ changes the sign by the nesting ${\mib Q}$ 
as shown in Fig. \ref{fig:FS}(e). 
Then, we can clarify the effect of the sign change by 
considering $\phi_{p_x}({\mib k})$.  

First, we calculate the dynamical charge susceptibility 
$\chi_{nn}({\mib q},\omega)$ in the superconducting state 
with the superconducting gap $\Delta_{\mib k}=\Delta(T) \phi({\mib k})$. 
This is given by 
\begin{eqnarray} &&
\chi_{nn}({\mib q},\omega)
=-\frac{1}{2}\sum_{\mib k} 
\left( 1+\frac{K_{{\mib k}+{\mib q}} K_{\mib k}
              -{\rm Re}\{ \Delta_{{\mib k}+{\mib q}}\Delta^\ast_{\mib k} \}}
{E_{{\mib k}+{\mib q}} E_{\mib k} } \right) 
\nonumber \\ && \times
\left( \frac{f(E_{{\mib k}+{\mib q}})-f(E_{\mib k})} 
            {E_{{\mib k}+{\mib q}}-E_{\mib k} -\omega} 
+ \frac{f(E_{{\mib k}+{\mib q}})-f(E_{\mib k})} 
            {E_{{\mib k}+{\mib q}}-E_{\mib k}+\omega} \right)
\nonumber \\ && 
-\frac{1}{2}\sum_{\mib k} 
\left( 1-\frac{K_{{\mib k}+{\mib q}} K_{\mib k}
              -{\rm Re}\{ \Delta_{{\mib k}+{\mib q}} \Delta^\ast_{\mib k} \}}
{E_{{\mib k}+{\mib q}} E_{\mib k} } \right) 
\nonumber \\ && \times
\left( \frac{f(E_{{\mib k}+{\mib q}})+f(E_{\mib k})-1} 
            {E_{{\mib k}+{\mib q}}+E_{\mib k}-\omega} 
+ \frac{f(E_{{\mib k}+{\mib q}})+f(E_{\mib k})-1} 
            {E_{{\mib k}+{\mib q}}+E_{\mib k}+\omega} \right) \qquad 
\label{eq:chinn}
\end{eqnarray} 
with $E_{\mib k}=(K_{\mib k}^2+|\Delta_{\mib k} |^2 )^{1/2}$ 
and the Fermi distribution function $f(E)$. 
We perform ${\mib k}$-summation within a Brillouin zone. 
The temperature dependence of $\Delta(T)$ is determined from the 
BCS relation, 
\begin{eqnarray} &&
1=-V_{\rm s} \sum_{\mib k} 
\frac{1-2f(E_{\mib k})}{ 2 E_{\mib k} } |\phi({\mib k})|^2 , 
\end{eqnarray}
where the pairing interaction $V_{\rm s}(<0)$ is determined by 
a given $\Delta(T=0)$. 

The neutron-scattering profile of a phonon is 
approximately the phonon spectral function $S({\mib q},\omega)$, i.e.,   
the imaginary part of the phonon Green's function, 
\begin{equation} 
S({\mib q},\omega)=-{\rm Im}D({\mib q},\omega). 
\end{equation}
The renormalized phonon Green's function  
$D({\mib q},\omega)$ is given by 
\begin{equation} 
D({\mib q},\omega)^{-1}=D_0({\mib q},\omega)^{-1}-\Pi({\mib q},\omega). 
\end{equation}
In the phonon selfenergy, polarizability is written as 
\begin{equation} 
\Pi({\mib q},\omega)=-|g_{\mib q}|^2 \chi_{nn}({\mib q},\omega)  
\end{equation}
in RPA.  
The bare phonon Green's function 
\begin{equation} 
D_0({\mib q})=\frac{2 \omega_{\mib q}}{\omega^2-\omega_{\mib q}^2}. 
\end{equation}
The electron-phonon coupling constant $g_{\mib q}$ and the phonon dispersion 
$\omega_{\mib q}$ are treated as constants, since we consider narrow 
${\mib q}$-region around ${\mib Q}$. 

\section{Phonon spectral function}
\label{sec:result}

\subsection{Normal state}

\begin{figure}[tbh]
\begin{center}
\includegraphics[width=8cm,height=7cm]{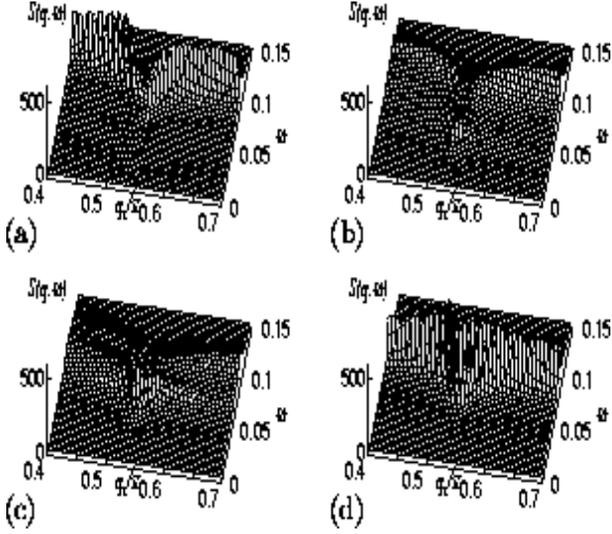}
\end{center}
\caption{\label{fig:sqws}
Phonon spectral function $S({\mib q}=(q_x,0,0),\omega)$ in the range 
$0.4 \le q_x/\pi \le 0.7$ for the normal state $\Delta=0$ (a) and the 
isotropic $s$-wave superconducting state with $\Delta=0.025$ (b), 
0.035 (c) and 0.050 (d). 
}
\end{figure}

In our calculation, we typically set $|g_{\mib q}|^2=0.1$, 
$\omega_{\mib q}=0.15$ and consider the low temperature case $T=0.001$, 
where $\Delta(T)\sim \Delta( T= 0)$. 
We present the phonon spectral function in the normal state 
$\Delta(T)=0$ in Fig. \ref{fig:sqws}(a). 
It shows  $S({\mib q},\omega)$ along the $q_x$-direction around ${\mib Q}$. 
By the effect of $\chi_{nn}({\mib q},\omega)$ in the normal state, 
the phonon dispersion is shifted to lower energy than $\omega_{\mib q}$, 
and it has a minimum at ${\mib Q}$ reflecting the nesting properties. 
The peak width of $S({\mib q},\omega)$ comes from 
${\rm Im}\chi_{nn}({\mib q},\omega)$. 
The temperature dependence of the phonon dispersion is weak 
in this parameter case. 
The minimum energy of the dispersion is $\omega=0.088$ at $T=0.05$. 
It slightly decreases to $\omega=0.086$ at $T=0.001$. 

When $\omega_{\mib q}$ is further decreased, 
the phonon dispersion at ${\mib q}\sim{\mib Q}$ approaches zero energy 
on lowering temperature by the Kohn anomaly. 
This phonon softening case is considered in \S\ref{sec:cdw}. 

\subsection{Isotropic $s$-wave superconducting state}

\begin{figure}[tbh]
\begin{center}
\includegraphics{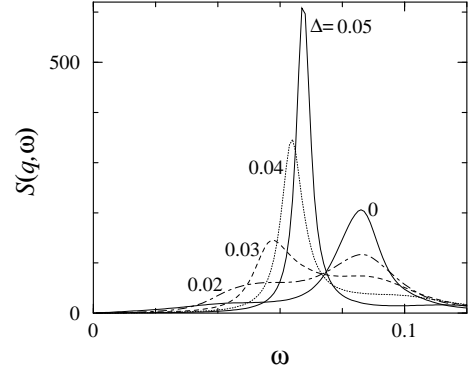}
\end{center}
\caption{\label{fig:wprfs}
Profiles of the phonon spectral function $S({\mib q},\omega)$ at 
${\mib q}=\pi(0.55,0,0)$ for $\Delta=0$, 0.02, 0.03, 0.04 and 0.05 
in the isotropic $s$-wave case. 
}
\end{figure}

We consider how $S({\mib q},\omega)$ is changed 
in the superconducting state. 
First, we investigate the isotropic $s$-wave case. 
We consider the $\Delta$-dependence of the 
phonon properties, where $\Delta \equiv \Delta(T=0)$. 
Figures \ref{fig:sqws}(b)-(d) show the change of $S({\mib q},\omega)$ 
with increasing $\Delta$. 
When $2 \Delta$ approaches the phonon dispersion from lower energy, 
the new peak of the gap mode appears below the phonon dispersion, 
as shown in Fig. \ref{fig:sqws}(b). 
The new peak is localized at $q_x \sim Q_x$, when it appears. 
With further increasing $2 \Delta$, the gap mode extends to wider 
$q_x$ region, as  shown in Fig. \ref{fig:sqws}(c).  
The peak position of the gap mode shifts to higher energy 
when $q_x$ is away from ${\mib Q}$. 
The original peak along the phonon dispersion is smeared above the gap mode. 
When $2\Delta$ becomes larger compared with the phonon dispersion, 
the original peak at the phonon dispersion vanishes, and the peak of 
the gap mode further grows up. 
To see this $\Delta$-dependence clearly, 
we show the $\omega$-profile of $S({\mib q},\omega)$ at 
$q_x=0.55 \pi$ in Fig. \ref{fig:wprfs}. 
The profiles at other $q_x$ also show the similar $\Delta$-dependence, 
while characteristic $\Delta$ value is changed. 
It is also noted that the lower energy tail of the peak is suppressed 
by increasing $\Delta$. 

\begin{figure}[tbh]
\begin{center}
\includegraphics{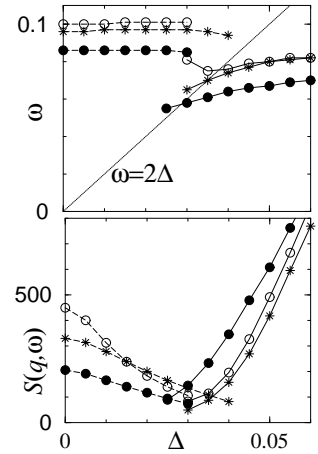}
\end{center}
\caption{\label{fig:peaks}
$\Delta$-dependence of the peak energy (upper panel) and the peak intensity 
(lower panel) for $q_x/\pi=0.50$ ($\circ$), 0.55 ($\bullet$), 0.60 ($\ast$).  
The points connected by solid lines  correspond to the peak of 
the gap mode.  
Dashed lines are for the peak of the original phonon dispersion. 
We also show the line $\omega=2\Delta$. 
}
\end{figure}

In Fig. \ref{fig:peaks}, we plot the $\Delta$-dependence of the peak 
energy and the peak intensity. 
We also show the cases $q_x=0.50 \pi$ and $0.60 \pi$ in addition to the 
case $q_x=0.55 \pi$ in order to see the $q_x$-dependence. 
The peak intensity of the original phonon dispersion at each $q_x$ is 
monotonically decreased with increasing $\Delta$, while the peak energy 
at the phonon dispersion is not changed. 
When $2 \Delta$ approaches the phonon dispersion, these original peaks 
are smeared, and the new peak of the gap mode appears at lower energy. 
It is seen as if the new peak first appears at finite $\omega$. 
If we could exclude any contributions of peak broadening, 
we might observe new peak at $\omega \sim 2 \Delta$ until the 
lower $\Delta$ case near $\Delta \sim 0$.\cite{Kee}   
However, this new peak is not observed in the lower $\Delta$ case, 
since the peak intensity is rapidly decreased on lowering $\Delta$ 
and this peak is completely smeared out by the effects of peak broadening. 
Then, new peak appears at finite energy in our results. 

Compared with the cases $q_x=0.5 \pi$ and $0.6 \pi$, 
the new peak appears from smaller 
$\Delta$ at $q_x=0.55 \pi$, since the phonon dispersion 
is located in lower energy at $q_x=0.55 \pi$. 
The increasing rate of the new peak's energy on raising $\Delta$ is 
much weaker than the relation $\omega\sim 2 \Delta$. 
At larger $\Delta$, the peak energy is apparently smaller than $2 \Delta$, 
and it seems to reduce to an $\omega$-independent value which is smaller 
than the original phonon dispersion. 
With increasing $\Delta$, the new peak becomes sharper and it has 
larger intensity. 
Along the dispersion for $q_x <Q_x$, both the original peak and the new 
peak are broad and overlapped when new peak appears, 
as shown in Fig. \ref{fig:sqws}(c). 
Because of the overlap, the energy of the new peak at $q_x=0.5 \pi$ in 
Fig. \ref{fig:peaks} is seen as if it appears at larger $\omega$ and 
decreases with increasing $\Delta$ at first. 

\begin{figure}[tbh]
\begin{center}
\includegraphics{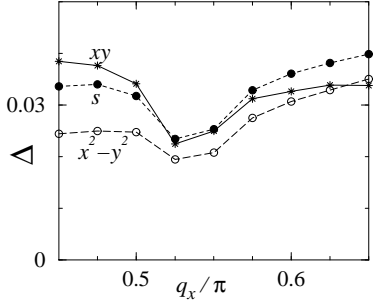}
\end{center}
\caption{\label{fig:cod}
$q_x$-dependence of the crossover $\Delta$,  
where peak intensity of the gap mode becomes larger 
than that of the original phonon peak, 
in the isotropic $s$-wave case ($\bullet$), the $s_{x^2-y^2}$ case 
($\circ$), and the $s_{xy}$ case ($\ast$).  
}
\end{figure}

In Fig. \ref{fig:cod}, we plot the $q_x$-dependence of the crossover 
$\Delta$, where the peak intensity of the gap mode becomes larger 
than that of the original phonon peak. 
Above the line of the figure at each $q_x$, the new peak of the gap 
mode is eminent. 
The crossover $\Delta$ is low near $q_x \sim Q_x$, and increases as 
$q_x$ is away from $Q_x$. 

The $\Delta$-dependence of the phonon profile in Fig. \ref{fig:wprfs} 
nicely reproduces the experimental data on 
YNi$_2$B$_2$C.\cite{Kawano,KawanoC,KawanoB} 
There, the $\Delta$-dependence is the temperature dependence $\Delta(T)$ 
or the magnetic field dependence $\Delta(H)$. 
In the experimental data, the energy of the new peak is close to $2 \Delta$, 
but the $T$-dependence of the peak energy $\omega$ is weaker than the 
relation $\omega = 2 \Delta (T)$. 
This $\Delta$-dependence of the peak energy is consistent with our results. 
The peak appears at the energy near $2 \Delta$, 
when $2 \Delta$ approaches the phonon dispersion. 
When $\Delta$ further increases, the increase of the peak energy is 
weaker than the relation  $\omega = 2 \Delta$, as shown in 
Fig. \ref{fig:peaks}. 
The experiment on LuNi$_2$B$_2$C reports that the phonon peak, 
which is shifted to lower energy, becomes sharp in the superconducting 
state.~\cite{Stassis} 
It is consistent to our result. 

\subsection{Anisotropic $s$-wave and $d_{x^2-y^2}$-wave 
superconducting states} 

In this subsection, we consider the anisotropic $s$-wave cases 
$s_{x^2-y^2}$ and $s_{xy}$ in order to clarify the effect of the 
superconducting gap suppression around the node points.  
While they have low energy excitations near nodes, 
we see that $S({\mib q},\omega)$ shows qualitatively same 
behavior as in the isotropic $s$-wave case   
presented in the previous subsection. 
That is, no drastic changes occur on phonon modes by their anisotropy. 
However, they show quantitatively different behaviors in the 
$\Delta$-dependence and in the $q_x$-dependence. 
We discuss these points in this subsection. 

The calculation of the $d_{x^2-y^2}$-wave case gives quantitatively 
almost the same $S({\mib q},\omega)$ as that of the $s_{x^2-y^2}$ case. 
This is because the nesting translation ${\mib Q}$ does not change the 
sign of the pairing function $\phi_{d_{x^2-y^2}}({\mib k})$ as shown in 
Fig. \ref{fig:FS}(c), and $|\phi_{d_{x^2-y^2}}({\mib k})|$ has the same 
gap amplitude as $|\phi_{s_{x^2-y^2}}({\mib k})|$.

\begin{figure}[tbh]
\begin{center}
\includegraphics[width=8cm,height=10.5cm]{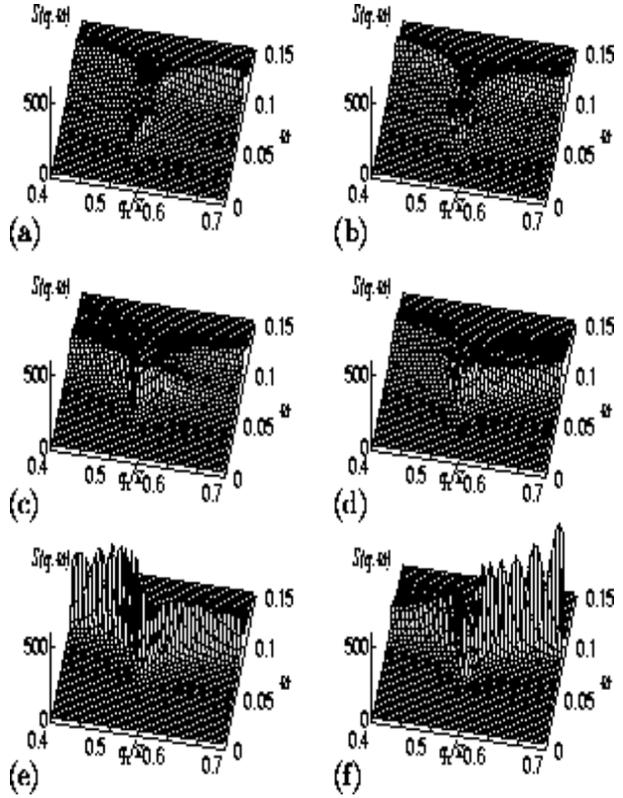}
\end{center}
\caption{\label{fig:sqwa}
Phonon spectral function $S({\mib q}=(q_x,0,0),\omega)$ in the range 
$0.4 \le q_x/\pi \le 0.7$ for the anisotropic $s$-wave cases.  
In the $s_{x^2-y^2}$ case, $\Delta=0.020$ (a), 0.030 (c) and 0.045 (e). 
In the $s_{xy}$ case, $\Delta=0.025$ (b), 0.035 (d) and 0.050 (f). 
The $d_{x^2-y^2}$-wave case gives the same structure as that of the 
$s_{x^2-y^2}$ case. 
}
\end{figure}

\begin{figure}[tbh]
\begin{center}
\includegraphics{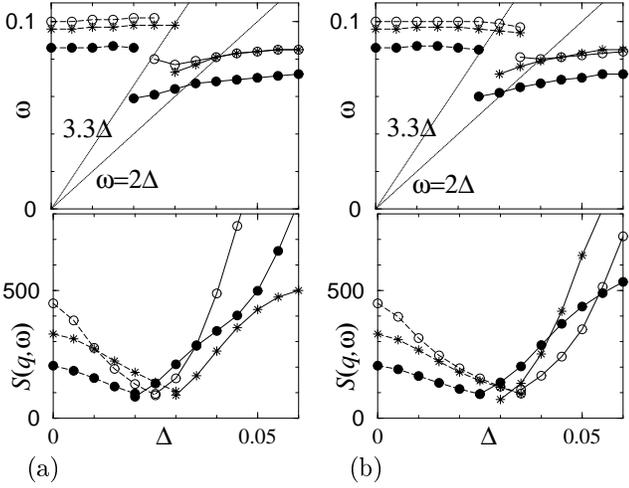}
\end{center}
\caption{\label{fig:peaka}
$\Delta$-dependence of the peak energy (upper panels) and the peak intensity 
(lower panels) for $q_x/\pi=0.50$ ($\circ$), 0.55 ($\bullet$), 0.60 ($\ast$) 
in the $s_{x^2-y^2}$ case (a) and in the $s_{xy}$-case (b). 
The points connected by solid lines correspond to the peak of the gap mode. 
Dashed lines are for the peak of the original phonon dispersion. 
We also show the lines $\omega =2 \Delta$ and $\omega =3.3 \Delta$.
}
\end{figure}

In Fig. \ref{fig:sqwa}, we present some typical behaviors of 
$S({\mib q},\omega)$ in the $s_{x^2-y^2}$ case and the $s_{xy}$ case. 
They show similar behavior as in the isotropic $s$-wave case 
in Fig.  \ref{fig:sqws}. 
When $2 \Delta$ approaches the minimum of the phonon dispersion, 
a new peak of the gap mode appears at ${\mib q}={\mib Q}$ [(a) and (b)]. 
With increasing $\Delta$, the new peak extends to the wide $q_x$-range 
[(c) and (d)]. 
When $\Delta$ further increases, the original peak at the phonon dispersion 
is completely smeared out, and the peak of the gap mode becomes sharper 
[(e) and (f)].  
The $\Delta$-dependence of the peak energy and intensity is 
plotted in Fig. \ref{fig:peaka}. 
The effect of the gap anisotropy appears at the onset $\Delta$ of the 
new peak and at the $q_x$-dependence of the peak intensity. 
As for the onset $\Delta$, we compare Figs. \ref{fig:peaks} and 
\ref{fig:peaka}. 
In the case of the anisotropic pairings, the superconducting 
gap is enhanced or suppressed by the factor $\phi({\mib k})$ in 
$\Delta_{\mib k}=\Delta(T)\phi({\mib k})$. 
In the $s_{x^2-y^2}$ case, $\phi_{s_{x^2-y^2}}({\mib k})$ is larger than 1 
along the flat Fermi surface $0 \le \theta < 30^\circ$ as seen in 
Fig. \ref{fig:FS}(b). At $\theta=0$, $\phi_{s_{x^2-y^2}}({\mib k})$ 
has maximum 1.65. 
Then, we also plot the line 
$\omega= 2 {\rm max}(\Delta_{\mib k})=3.3\Delta$  in addition to the 
line $\omega= 2 \Delta$ in Fig. \ref{fig:peaka}. 
In Fig. \ref{fig:peaka}(a), the onset $\Delta$ of the new peak is lower 
than that of the isotropic $s$-wave case, but it is larger than the 
relation $\omega= 3.3 \Delta$. 
That is, since the effect of $\phi_{s_{x^2-y^2}}({\mib k})$ is averaged on 
the flat Fermi surface, the enhancement factor is larger than 1, but 
smaller than 1.65. 
Compared with the $s_{x^2-y^2}$ case, the onset $\Delta$ is larger in 
the $s_{xy}$ case, since the average of the factor 
$\phi_{s_{xy}}({\mib k})$ on the flat Fermi surface is smaller. 

As for the $q_x$-dependence of the peak intensity, 
the intensity of the gap mode in Fig. \ref{fig:sqwa} is enhanced for 
$q_x < Q_x$ ($q_x > Q_x$) in the $s_{x^2-y^2}$ case ($s_{xy}$ case) 
compared with the isotropic $s$-wave case. 
The gap mode eminently extends toward $q_x < Q_x$ 
($q_x > Q_x$)  in Fig. \ref{fig:sqwa}(c) (Fig. \ref{fig:sqwa}(d)).  
These characteristics are also recognized in Fig. \ref{fig:peaka}, 
where the peak intensity of the gap mode at $q_x=0.5 \pi$ ($0.6 \pi$) 
is largely enhanced, and in Fig. \ref{fig:cod}, where crossover $\Delta$ 
is lower for $q_x < Q_x$ ($q_x > Q_x$).  
This difference comes from the ${\mib k}$-dependence of $\phi({\mib k})$ 
along the perpendicular direction to the flat Fermi surface. 
We explain it by the $k_x$-dependence near the flat Fermi surface 
at $k_x=0.275\pi$ in Fig. \ref{fig:FS}(a). 
In the $s_{x^2-y^2}$ case, the factor $\phi_{s_{x^2-y^2}}({\mib k})$ 
is enhanced when $k_x$ is decreased because of the term $\cos k_x$ 
in $\phi_{s_{x^2-y^2}}({\mib k})$. 
Then, for the smaller $q_x (\sim 2 k_x)$ than the nesting vector $Q_x$, 
the superconducting gap is enhanced, and the intensity of the gap mode 
becomes eminent. 
On the other hand, $\phi_{s_{xy}}({\mib k})$  is enhanced when $k_x$ is 
increased because of the factor $\sin k_x$ in $\phi_{s_{xy}}({\mib k})$. 
Then, for larger $q_x (\sim 2 k_x)$ than $Q_x$, the gap mode is eminent. 

\subsection{$d_{xy}$-wave and $p_{x}$-wave superconducting states} 

In this subsection, we examine the effect coming from the sign 
change of the pairing function by the nesting translation.  
The $d_{xy}$-wave case has the same superconducting gap as that 
of the $s_{xy}$ case. 
But, the sign of the pairing function $\phi_{d_{xy}}({\mib k})$ 
is changed by the nesting translation ${\mib Q}$ as shown in 
Fig. \ref{fig:FS}(d). 
Then, the phonon behavior in the  $d_{xy}$-wave case is qualitatively 
different from the $s_{xy}$ case. 
It is because ${\rm Re}\{ \Delta_{{\mib k}+{\mib q}}\Delta_{\mib k}^\ast \}$ 
of $\chi_{nn}({\mib q},\omega)$ in eq. (\ref{eq:chinn}) becomes negative 
at ${\mib q}\sim{\mib Q}$ when the sign change occurs.~\cite{Kee2}  
The $p_x$-wave case also shows the same behavior due to the sign change 
of $\phi_{p_x}({\mib k})$ through the nesting translation. 

\begin{figure}[tbh]
\begin{center}
\includegraphics[width=8cm,height=3.5cm]{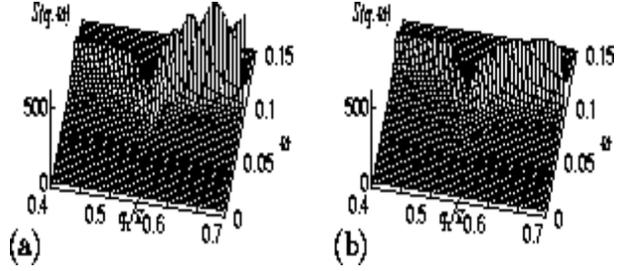}
\end{center}
\caption{\label{fig:sqwd}
Phonon spectral function $S({\mib q}=(q_x,0,0),\omega)$ in the range 
$0.4 \le q_x/\pi \le 0.7$ 
for the $d_{xy}$-wave case (a) and the $p_{x}$-wave case (b). 
$\Delta=0.030$.
}
\end{figure}

\begin{figure}[tbh]
\begin{center}
\includegraphics{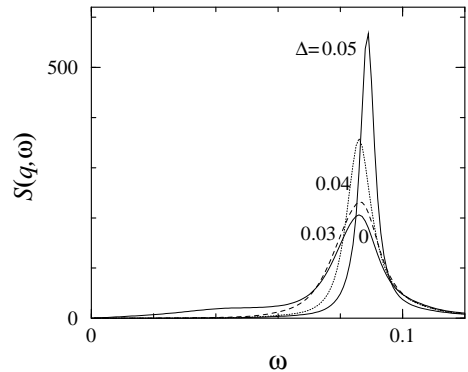}
\end{center}
\caption{\label{fig:wprfd}
Profiles of the phonon spectral function $S({\mib q},\omega)$ at 
${\mib q}=\pi(0.55,0,0)$ for $\Delta=0$, 0.03, 0.04 and 0.05 
in the $d_{xy}$-wave case. 
}
\end{figure}

\begin{figure}[tbh]
\begin{center}
\includegraphics{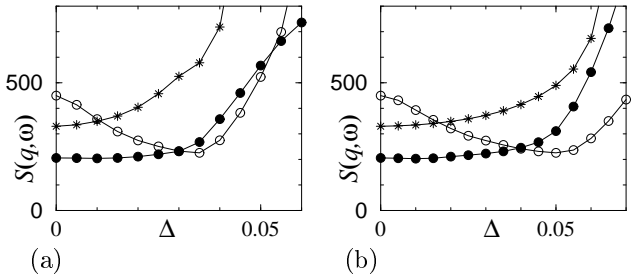}
\end{center}
\caption{\label{fig:peakd}
$\Delta$-dependence of the peak intensity for $q_x/\pi=0.50$ ($\circ$), 
0.55 ($\bullet$), 0.60 ($\ast$) in the $d_{xy}$-wave case (a) 
and the $p_x$-wave case (b).  
}
\end{figure}

Figure \ref{fig:sqwd} shows the typical behavior of $S({\mib q},\omega)$ 
in the $d_{xy}$-wave (a) and the $p_x$-wave (b) cases. 
There, the new peak of the gap mode does not appear. 
By the effect of $\Delta$, the peak intensity of the original phonon mode 
is modified. 
But, the peak energy along the phonon dispersion is almost unchanged. 
To see this $\Delta$-dependence, we show the $\omega$-profile of 
$S({\mib q},\omega)$ at $q_x=0.55 \pi$ for the $d_{xy}$-wave case 
in Fig. \ref{fig:wprfd}. 
With increasing $\Delta$, the low energy tail is gradually suppressed. 
After $2\Delta$ exceeds the phonon dispersion, 
the peak intensity becomes sharper and the intensity is enhanced with 
increasing $\Delta$. 
To see the $q_x$-dependence of this behavior, we plot the 
$\Delta$-dependence of the peak intensity at $q_x/\pi=0.50$, 0.55, 0.60 
in Fig. \ref{fig:peakd}. 
At low $\Delta$, the peak intensity at  $q_x=0.50 \pi$ ($0.60 \pi$) 
is suppressed (enhanced) with increasing $\Delta$, as also shown in 
Fig. \ref{fig:sqwd}. 
This $q_x$-dependence comes from the $k_x$-dependence of the pairing 
functions $|\phi_{d_{xy}}({\mib k})|$ and $|\phi_{p_{x}}({\mib k})|$ 
near the flat Fermi surface at $k_x=0.275\pi$. 
There, $|\phi_{d_{xy}}({\mib k})|$ and $|\phi_{p_{x}}({\mib k})|$ is 
enhanced with increasing $k_x$ because of the factor $\sin k_x$ 
in the pairing functions.  
Then, for larger $q_x (\sim 2 k_x)$ than $Q_x$, 
the superconducting gap is stronger, and the peak intensity is enhanced. 


\section{Suppression of phonon softening} 
\label{sec:cdw}

When $\omega_{\mib q}$ is enough low, the phonon dispersion approaches 
zero energy at ${\mib q}={\mib Q}$ on lowering 
temperature due to the Kohn anomaly.  
This phonon softening causes the second order transition to the 
CDW states. 
It is suggested that this softening is suppressed by the 
onset of superconductivity.\cite{Kee}  
To study this phenomenon, we consider the phonon properties in the case 
$\omega_{\mib q}=0.117$. 
The other parameters are unchanged. 

\begin{figure}[tbh]
\begin{center}
\includegraphics[width=8cm,height=7cm]{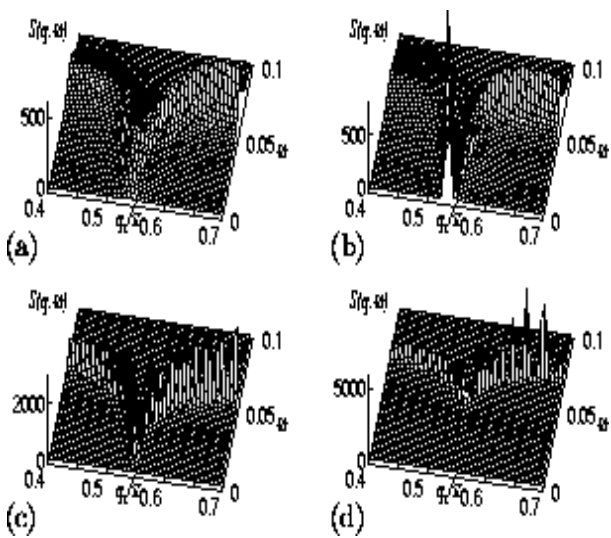}
\end{center}
\caption{\label{fig:sqwL}
Phonon spectral function $S({\mib q}=(q_x,0,0),\omega)$ in the range 
$0.4 \le q_x/\pi \le 0.7$ for $\omega_{\mib q}=0.117$. 
In the normal state($\Delta=0$), $T=0.050$ (a) and $0.005$ (b). 
In the superconducting state of the $s_{xy}$-case (c) and 
the $d_{xy}$-wave case (d), $\Delta=0.05$ and $T=0.005$.  
}
\end{figure}

Figure \ref{fig:sqwL} shows $S({\mib q},\omega)$ in this case. 
At high temperature $T=0.050$ (a), the phonon dispersion is located 
at finite energy. 
With decreasing temperature, the phonon dispersion is shifted to lower 
energy within the narrow region at ${\mib q}={\mib Q}$. 
At low temperature $T=0.005$ (b) in the normal state, 
the minimum of the dispersion touches zero energy, suggesting CDW transition. 
However, in the superconducting state, the phonon dispersion remains 
at finite energy, meaning the suppression of the CDW transition. 
Figure \ref{fig:sqwL}(c) is for the $s_{xy}$ case. 
We obtain this type phonon behavior in the isotropic and anisotropic 
$s$-wave cases, and the $d_{x^2-y^2}$-wave case, i.e., 
in the case when the sign of the pairing function is not changed 
by the nesting translation. 
When the sign is changed by the nesting, such as the case $d_{xy}$-wave 
and $p_x$-wave cases, 
the phonon dispersion is further shifted to higher energy. 
We show $S({\mib q},\omega)$ of the $d_{xy}$-wave case in 
Fig. \ref{fig:sqwL}(d). 
It is noted that these change of the phonon dispersion occurs 
only near ${\mib q}={\mib Q}$. 
The dispersion is not largely changed far away from ${\mib Q}$, e.g., 
at $q_x=0.4 \pi$ or $0.7 \pi$ in Fig. \ref{fig:sqwL}. 

\begin{figure}[tbh]
\begin{center}
\includegraphics{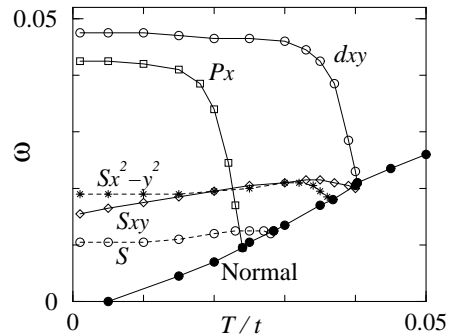}
\end{center}
\caption{\label{fig:peakL}
Temperature dependence of the minimum energy of the phonon dispersion 
for $\omega_{\mib q}=0.117$ in the normal state ($\Delta=0$) and the 
superconducting state with $\Delta(T=0)=0.05$. 
In the superconducting state, we present the isotropic $s$-wave, 
the anisotropic $s$-wave $s_{x^2-y^2}$ and $s_{xy}$, 
the $d_{xy}$-wave and the $p_x$-wave cases. 
The points for the $d_{x^2-y^2}$-wave is the same as the $s_{x^2-y^2}$ case. 
}
\end{figure}

We plot the temperature dependence of the minimum energy of the 
phonon dispersion in Fig. \ref{fig:peakL}, where the phonon dispersion 
is identified by the peak energy. 
In the normal state,  the minimum energy $\omega$ monotonically decreases 
on lowering temperature. 
At $T \sim 0.005$, the CDW transition occurs by the softening 
$\omega \rightarrow 0$. 
On the other hand, the softening stops in the superconducting state. 
Then, the second order transition to the CDW state does not occur. 
The minimum $\omega$ shows weak temperature dependence below $T_c$ 
in the $s$-wave cases and the $d_{x^2-y^2}$-wave case. 
However, the minimum $\omega$ is largely enhanced just below $T_c$ 
and it saturates at low temperature in the $d_{xy}$-wave 
and the $p_x$-wave cases, i.e., in the case when the pairing function 
changes the sign through the nesting translation. 
In Fig. \ref{fig:peakL}, $T_c$ is different due to the factor 
$\phi({\mib k})$, while we set $\Delta(T=0)=0.05$ in each pairing case. 

The amplitude of the superconducting gap is also controlled by the external 
magnetic field. 
In the case when the CDW transition is suppressed by superconductivity, 
the CDW state is recovered if we destroy the superconductivity by 
applying the magnetic field. 
This prediction is testable for appropriate superconductors, 
possibly YNi$_2$B$_2$C. 

\section{Summary and discussions}
\label{sec:summary}

By extending the previous theories~\cite{Balseiro,Littlewood,Allen,Kee} 
on the appearance of the new phonon mode below $T_{\rm c}$ for isotropic 
$s$-wave superconductors, we have investigated the detailed features 
of the dispersion relation for this new phonon gap mode in light of 
various possible anisotropic pairings. 
It is demonstrated that observation of this dispersive gap mode can be 
a useful spectroscopic tool for sorting out some of the possible 
pairing states.  
With increasing the amplitude of the superconducting gap $\Delta$, 
this new peak of the gap mode becomes sharp and the original phonon 
peak is smeared. 
We have estimated the $\Delta$-dependence of the dispersion and the 
intensity of the peak. 
The effect of the gap anisotropy is classified to two cases; 
(i) the case when the sign of the pairing function is not changed 
through the nesting translation, and 
(ii) the case when the sign is changed. 
In the former case (i), the gap anisotropy modifies the phonon spectral 
function quantitatively. 
The anisotropy effects appear at the onset $\Delta$ of the gap mode 
and at the $q_x$-dependence of the peak intensity, 
reflecting local amplitude of the superconducting gap 
at the Fermi surface region related to the nesting translation. 
In the latter case (ii), the new peak of the gap mode does not appear. 
There, for larger $\Delta$, the original phonon peak becomes sharp 
without large shift of the peak energy. 
While we obtain these properties by analyzing example models 
for the Fermi surface and pairing functions, 
we can expect that these qualitative results 
coming from the superconducting gap anisotropy are independent of the model, 
because of the generality of their origin. 

In the experimental data on YNi$_2$B$_2$C, 
the eminent dispersive new peak of the gap mode is observed around 
${\mib q}={\mib Q}$.~\cite{Kawano,KawanoC,KawanoB} 
This behavior of the gap mode is qualitatively consistent to our 
numerical results. 
The phonon peak on LuNi$_2$B$_2$C is shifted to lower energy 
in the superconducting state.\cite{Stassis}  
Then, in these materials, we can exclude the pairing symmetry of 
the case (ii), for example, $d_{xy}$-wave. 
That is, we can conclude that the pairing function has the same sign 
at the flat Fermi surface related to the nesting.  

Further identification of the pairing symmetry among isotropic $s$-wave, 
anisotropic $s$-wave and $d_{x^2-y^2}$-wave is difficult, 
because it needs quantitative considerations. 
But, we give some discussions here. 
One of the key points to identify the pairing symmetry is 
the $q_x$-dependence of the peak intensity. 
In the data on YNi$_2$B$_2$C, the peak intensity is larger 
for $q_x < Q_x$.~\cite{Kawano}  
Following our analysis, this suggests that the gap anisotropy of the type 
$|\cos k_x -\cos k_y |$ is preferable. 
The second point is that the superconducting gap largely opens at the 
flat Fermi surface related to the nesting, when the eminent gap mode 
appears. 
The contribution of the superconducting gap affects on the phonon as an 
average along the flat Fermi surface. 
In our calculation, the average of $\phi_{s_{xy}}({\mib k})$ is smaller than 
that of $\phi_{s_{x^2-y^2}}({\mib k})$. 
The former has a node at the flat part of the Fermi surface. 
Since the flat part is long in our model of the Fermi surface, 
the both averages do not show large difference. 
However, if the flat part is shorter, the average of the gap is smaller 
in the case when the node of the gap anisotropy is located at the flat part. 
In this case, we cannot expect the eminent gap mode. 
To obtain decisive conclusion on these above-mentioned points, 
we need to establish the knowledge about the detailed shape of the 
Fermi surface and the nesting properties. 

Note added: 
Recently, a similar neutron experiment has been reported on 
${\rm ErNi_2B_2C}$ by Kawano-Furukawa {\it et al.}, 
showing the gap mode below $T_{\rm c}$.\cite{KawanoEr} 

\section*{Acknowledgments}
We would like to thank H. Yoshizawa for helpful discussions.

%

\end{document}